\documentclass[conference]{IEEEtran}
\IEEEoverridecommandlockouts
\usepackage[T1]{fontenc}
\usepackage{cite}
\usepackage{amsmath,amssymb,amsfonts}
\usepackage{algorithmic}
\usepackage{graphicx}
\usepackage{textcomp}
\usepackage{xcolor}
\usepackage{pgfplots}
\pgfplotsset{compat=1.18}
\usepackage{siunitx}
\usepackage{url}
\pgfplotsset{empty legend/.style={legend image code/.code={}}}
\def\BibTeX{{\rm B\kern-.05em{\sc i\kern-.025em b}\kern-.08em
    T\kern-.1667em\lower.7ex\hbox{E}\kern-.125emX}}
\begin{document}

\title{Machine Learning for Scalable Multi-AP Coordination in IEEE 802.11bn Networks
\thanks{This research was funded by the National Science Centre, Poland (DEC-2020/39/I/ST7/01457).}
}

\author{
\IEEEauthorblockN{Maksymilian Wojnar}
\IEEEauthorblockA{\textit{AGH University of Krakow} \\
Krakow, Poland \\
maksymilian.wojnar@agh.edu.pl\vspace{-11.7pt}}
\and
\IEEEauthorblockN{Szymon Szott}
\IEEEauthorblockA{\textit{AGH University of Krakow} \\
Krakow, Poland \\
szymon.szott@agh.edu.pl\vspace{-11.7pt}}
}

\maketitle
\bstctlcite{IEEEexample:BSTcontrol}

\begin{abstract}
As Wi-Fi deployments become denser, fully distributed random channel access no longer guarantees the required quality of service. The multi-access point coordination (MAPC) framework introduced in IEEE~802.11bn lets neighboring access points (APs) cooperate, but choosing which APs transmit together and how to configure them remains an open, computationally hard problem. We develop and validate a stack of machine learning (ML) coordination methods, progressing from lightweight hierarchical multi-armed bandits (H-MABs) to a generative pipeline built on graph neural networks (GNNs) and flow matching (FM). The methods are evaluated in simulation and on a hardware testbed, improving throughput over legacy access and scaling coordination to dozens of APs where analytical and search-based methods become intractable.
\end{abstract}

\begin{IEEEkeywords}
IEEE 802.11bn; coordinated spatial reuse; flow matching; multi-armed bandits; neural networks; reinforcement learning.
\end{IEEEkeywords}

\section{Introduction}
The increasing density of Wi-Fi networks exposes the limits of fully distributed, contention-based channel access, which cannot ensure high and predictable service quality. In response, IEEE~802.11bn (Wi-Fi~8) introduces multi-access point coordination (MAPC)~\cite{galatigiordano2024wifi8}, letting neighboring access points (APs) share radio resources more efficiently. Among MAPC variants, our research targets coordinated spatial reuse (Co-SR), in which selected nodes transmit simultaneously on the same channel with carefully chosen transmit power, so that transmissions are successfully decoded while avoiding mutual interference. MAPC currently targets downlink transmissions only, and the standard enforces just the basic signaling: it leaves open the selection of the transmitting AP--station pairs and their transmit power. The modulation and coding scheme (MCS), although outside the standard, is another major component that determines the final performance. We call the joint selection of these parameters as the \emph{Co-SR scheduling problem}, the central problem of our work. Crucially, the current IEEE~802.11bn specification limits Co-SR to pairs of APs, as coordinating quickly becomes prohibitively complex. Nevertheless, all the coordination methods we propose go beyond Wi-Fi~8, where any number of APs may cooperate. Such network-wide coordination is most valuable in large, centrally managed deployments, such as industrial, corporate, and institutional networks.

The problem is computationally hard and scales poorly: the number of possible configurations grows exponentially with the number of nodes in a network, and parameter selection must account for variable, hard-to-predict propagation. Existing Co-SR solutions fall into a few classes, each with a shared weakness: (i) measurement-driven heuristics and centralized schemes~\cite{haxhibeqiri2024centralized,nunez2023group}, bound by costly signaling and unreliable received signal strength (RSS) measurements; (ii) iterative solver-based and adaptive coordination~\cite{zhu2025enhanced}, bound by poor scaling; (iii) and learning-based schedulers~\cite{jung2026shared}, so far confined to small static downlink topologies. None establishes the theoretical performance limits of Co-SR or offers scalable machine learning (ML) methods verified in complex scenarios. Addressing these gaps, we study lightweight bandit-based schedulers that learn online, analytical performance bounds for Co-SR, and a generative one-shot pipeline that scales to dozens of APs. Our methods are available as open source\footnote{\url{https://github.com/ml4wifi-devs/csr}}\textsuperscript{,}\footnote{\url{https://github.com/ml4wifi-devs/fm4wifi}}.

\section{Methodology}
Established simulators, such as ns-3, lack MAPC support. Therefore, we build an open-source Monte Carlo Co-SR simulator~\cite{wojnar2025hmab}, which computes the interference between nodes and the success probability of reception, thereby approximating the throughput achieved under parallel transmissions. This effective data rate, summed over all parallel transmissions, serves as the optimization objective. We first formulate Co-SR scheduling as a bandit problem and develop a hierarchical multi-armed bandit (H-MAB) whose levels select the transmitting APs, their receiving stations, and transmit power, with optional clustering to make the action space tractable in dense topologies~\cite{wojnar2025hmab}. To quantify achievable performance, we formulate mixed-integer linear programs (MILP) optimizing for throughput (T-Optimal) or fairness (F-Optimal)~\cite{wojnar2025jsac}. Finally, we represent MAPC topologies and configurations as graphs and encode them with a graph neural network (GNN) variational autoencoder~\cite{wojnar2025gnn}. A flow matching (FM) model, chosen for its high-quality few-step generation, then produces new configurations, and a surrogate rate predictor scores them without costly simulator calls~\cite{wojnar2026fm4wifi}. Validation is threefold: against the analytical MILP upper bounds, in simulation (versus legacy DCF and IEEE 802.11ax spatial reuse) across static and dynamic topologies, and on a real openwifi testbed~\cite{jiao2020openwifi}, whose digital twin serves for preliminary hyperparameter tuning.

\section{Selected Results}
We first model Co-SR group selection as a bandit problem and solve it with H-MAB~\cite{wojnar2025hmab}. Within seconds, the agents learn which AP--station pairs perform best in a given topology and adapt to changes such as station mobility. The hierarchy outperforms a flat MAB by rejecting incompatible APs at a higher decision level, and UCB emerges as the best general-purpose underlying algorithm. Extending H-MAB with transmit-power selection~\cite{wojnar2025jsac} raises aggregate throughput over DCF by 80\% on average in random scenarios, without reducing per-station transmission opportunities, and outperforms 802.11ax spatial reuse. For the densest office topologies, clustering the network drastically shortens convergence time. On a real openwifi testbed~\cite{jiao2020openwifi}, the agents adapt to real radio conditions.

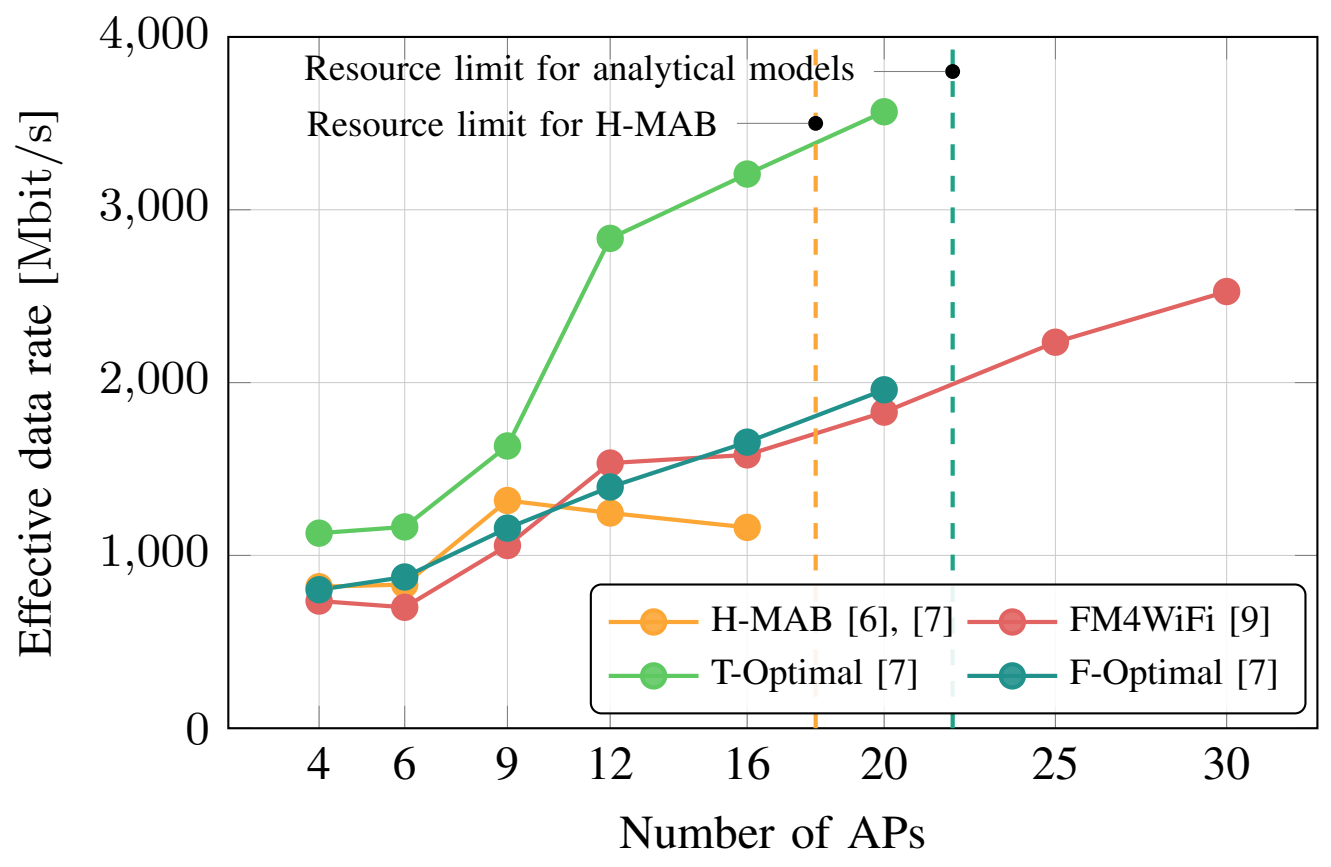
\begin{figure}[t]
\centering
\resizebox{\columnwidth}{!}{\begin{tikzpicture}
\definecolor{palA}{RGB}{58,82,139}
\definecolor{palB}{RGB}{33,145,140}
\definecolor{palC}{RGB}{31,163,134}
\definecolor{palD}{RGB}{94,201,97}
\definecolor{palE}{RGB}{204,71,20}
\definecolor{palF}{RGB}{225,100,98}
\definecolor{palG}{RGB}{242,132,75}
\definecolor{palH}{RGB}{252,166,54}
\colorlet{ctopt}{palD}
\colorlet{cfopt}{palB}
\colorlet{chmab}{palH}
\colorlet{cfm}{palF}
 \tikzset{
        every pin/.style={rectangle,rounded corners=5pt,font=\scriptsize},
        small dot/.style={fill=black,circle,scale=0.3},
    }
\begin{axis}[
  width=\columnwidth,
  height=0.7\columnwidth,
  xlabel={Number of APs},
  ylabel={Effective data rate [\si{\mega\bit\per\second}]},
  ymin=0, ymax=4000,
  ytick distance=1000,
  xtick={3.5,6,9,12,16,20,25,30},
  xticklabels={4,6,9,12,16,20,25,30},
  grid=both,
  major grid style={line width=0.2pt, draw=gray!40},
  minor grid style={line width=0.1pt, draw=gray!20},
  enlarge x limits=0.1,
  clip=false,
  tick label style={font=\small},
  xticklabel style={font=\small, align=center},
  label style={font=\small},
  every axis plot/.append style={thick, mark size=2.2pt},
  legend style={
    at={(0.99,0.02)},
    anchor=south east,
    draw=black,
    fill=white,
    fill opacity=0.9,
    text opacity=1,
    font=\scriptsize,
    rounded corners=2pt
  },
  legend cell align=left,
  legend columns=2
]
\addplot[color=chmab, solid, mark=*, mark options={solid}] coordinates {(3.5,819.045952) (6,831.202042) (9,1317.857112) (12,1245.538993) (16,1162.109409)};
\addlegendentry{H-MAB~\cite{wojnar2025hmab,wojnar2025jsac}}

\addplot[color=cfm, solid, mark=*, mark options={solid}] coordinates {(3.5,736.0674234135665) (6,699.8786241794309) (9,1057.195363785558) (12,1534.3886761487965) (16,1581.8278172866521) (20,1829.1096827133479) (25,2233.6313594091903) (30,2526.9471416849015)};
\addlegendentry{FM4WiFi~\cite{wojnar2026fm4wifi}}

\addplot[color=ctopt, solid, mark=*, mark options={solid}] coordinates {(3.5,1128.9) (6,1165.0) (9,1633.4) (12,2834.4) (16,3206.7) (20,3567.1)};
\addlegendentry{T-Optimal~\cite{wojnar2025jsac}}

\addplot[color=cfopt, solid, mark=*, mark options={solid}] coordinates {(3.5,801.5038355813325) (6,875.2661630544156) (9,1158.0232544994062) (12,1395.8932573287711) (16,1655.4317356008007) (20,1958.243962961537)};
\addlegendentry{F-Optimal~\cite{wojnar2025jsac}}

\addplot[dashed, thick, color=chmab] coordinates {(18,0) (18,4000)};
\node [small dot,pin=-180:{Resource limit for H-MAB}]    at (axis cs:18,3500)   {};

\addplot[dashed, thick, color=palC] coordinates {(22,0) (22,4000)};
\node [small dot,pin=-180:{Resource limit for analytical models}]  at (axis cs:22,3800)   {};

\end{axis}
\end{tikzpicture}}
\caption{Effective data rate versus number of APs in a static residential (office-grid) scenario, with every method selecting APs, AP--station pairs, and transmit power under ideal MCS.}
\label{fig:static_residential}
\end{figure}

We then study representation learning for MAPC configurations~\cite{wojnar2025gnn}, modeling the network as a heterogeneous graph and training a GNN variational autoencoder whose loss accounts for missing attributes. Discrete attributes such as MCS and transmission power are reconstructed with near 100\% accuracy, while continuous ones (RSS and success probability) are inherently harder, reaching 15--50\% accuracy. Moreover, the embeddings are semantically organized, therefore, we use them as a base for our FM4WiFi generative pipeline~\cite{wojnar2026fm4wifi}, which reformulates Co-SR scheduling as a conditional generation problem. The pipeline consists of a GNN autoencoder, an FM sampler, and a surrogate rate predictor. FM4WiFi jointly optimizes the selection of transmitters (APs), receivers (stations), MCS and transmit power levels. It produces a schedule in a single inference step in under one second and scales beyond 30 APs, where analytical and reinforcement learning (RL) methods become infeasible. Since only FM4WiFi jointly selects the MCS, Fig.~\ref{fig:static_residential} gives each method an ideal MCS for a fair comparison. FM4WiFi matches or exceeds the effective data rate of the strongest baselines at medium-to-large scale, and keeps improving past the network sizes where H-MAB and the analytical bounds hit their resource limits. In dynamic scenarios, FM4WiFi increases the effective data rate by roughly 25\% over a static configuration with regeneration triggered by network events or at cyclic intervals.

H-MAB and FM4WiFi complement each other, and both coordinate much more than two APs as in Wi-Fi~8. H-MAB is lightweight and well suited to small and medium networks, while the FM4WiFi provides the sub-second decisions and scalability needed for large topologies. The T-Optimal and F-Optimal upper bounds serve as analytical reference for both. Beyond the referenced publications, the H-MAB and FM4WiFi results have been presented to the IEEE~802.11 AIML Topic Interest Group~\cite{szott2025tigmab,szott2026tigml}, supporting ongoing Wi-Fi standardization efforts.

\section{Discussion and Future Work}
Overall, this work establishes ML as a practical and scalable foundation for Co-SR coordination beyond Wi-Fi~8, combining lightweight online agents, analytical performance bounds, and a scalable generative pipeline validated through analysis, simulation, and hardware. Several directions remain open. We plan to add RL to the FM4WiFi pipeline to further increase performance and enable online adaptation at a large scale. We are also developing approximate performance bounds that scale better than the MILP formulation. Finally, we are exploring optimization driven directly by the surrogate model.

\bibliographystyle{IEEEtran}
\bibliography{bibliography}

\end{document}